\documentclass{aa}
\def\sax{{\it BeppoSAX\,\,}}
\def\lae{\mathrel{<\kern-1.0em\lower0.9ex\hbox{$\sim$}}}
\def\gae{\mathrel{>\kern-1.0em\lower0.9ex\hbox{$\sim$}}}

\usepackage{epsfig}
\begin{document}
\title{\sax Spectral Survey of BL Lacs - new spectra and results}
\author{V. Beckmann\inst{1,}\inst{2}, A. Wolter\inst{3},
  A. Celotti\inst{4}, L. Costamante\inst{3}, G. Ghisellini\inst{3},
  T. Maccacaro\inst{3}, and G. Tagliaferri\inst{3}}
\offprints{Volker.Beckmann@obs.unige.ch}
\institute{INTEGRAL Science Data Centre, Chemin d' Ecogia 16, CH-1290
 Versoix, Switzerland
 \and Institut f\"ur Astronomie und Astrophysik, Universit\"at T\"ubingen, Sand 1, D-72076 T\"ubingen, Germany
 \and Osservatorio Astronomico di Brera, Via Brera 28, I-20121 Milano,
 Italy
 \and SISSA/ISAS, Via Beirut 2-4, I-34014 Trieste, Italy}
\date{Received date; accepted date}
\authorrunning{Beckmann et al.}
\titlerunning{\sax Spectral Survey of BL Lacs}

\abstract{ We present \sax LECS, MECS, and PDS spectra of eleven X-ray
selected BL Lacertae objects.\linebreak Combining these sources with
the ones presented elsewhere we have a sample of 21 BL Lacs from the
Einstein Medium Sensitivity and Einstein Slew Survey. The sample shows
strong correlations of several physical parameters with the peak
frequency of the synchrotron branch of the spectral energy
distribution. In particular the peak frequency is correlated to the
X-ray spectral shape: objects with the peak near to the X-ray band
show harder and straighter X-ray spectra than those of the low energy
peaked sources.
This work shows that the recently proposed unification scenario for
different types of blazars can hold also within the class of high
frequency peaked BL Lac objects.
 }

\maketitle

\section{Introduction}
BL Lac objects mark one extreme population of active galactic nuclei
(AGN). They exhibit intense variability (e.g. Wagner \& Witzel
\cite{variability}) and polarization (e.g. K\"uhr \& Schmidt
\cite{polarization}) but do not show strong emission or absorption
lines in their spectrum (e.g. Stocke et al. \cite{stocke}).  In order
to understand their physical nature through the emission processes
involved and their relative contribution, one has to study their
spectral energy distribution (SED).

Determining the SED is a telescope time consuming issue. There exist
some well studied bright BL Lac objects - like Mkn 421, 1ES 2344+512
and Mkn 501 - for which detailed measurements at several wavelengths
are available, but to construct a global scenario for BL Lac objects
it is necessary to also derive the properties of complete
samples. This means investigating fainter objects too.  Because of its
sensitivity and wide energy band, which allows us to study
simultaneously the BL Lac spectrum from 0.1 to 100 keV, the \sax
satellite is a powerful tool for this aim.

Here we present a well defined sample of 21 BL Lac objects which have
been studied using \sax.  In the following Section 2 we define the
sample; observations and data analysis are explained in Section 3. The
X-ray data are compared to earlier results by the ROSAT satellite in
Section 4, and in Section 5 we examine the properties of the BL Lac
SEDs. The results are discussed in the context of unified schemes for
BL Lac objects in Section 6.

\section{The sample}

Because strong emission lines are missing in the optical spectra of BL
Lac objects, this kind of AGN is usually selected either in the radio
or in the X-ray band. The X-ray selected BL Lacs are mainly found in
the surveys carried out by the {\em EINSTEIN IPC} (the EMSS Sample, Gioia et
al. \cite{emss}, and the Slew Survey Sample, Perlman et al. \cite{ess}), 
and by the {\em ROSAT PSPC} (ROSAT All-Sky Survey
Sample; Bade et al. \cite{bade}, Beckmann \cite{thesis}). The Slew
Survey Sample covers the whole high Galactic latitude sky, while the
EMSS is deeper, but only over an area of $\sim 800 \, \rm
deg^2$. By selecting objects with fluxes $F_X(0.1 - 10 \rm \, keV) \ge
10^{-11} erg \, cm^{-2} \, sec^{-1}$ in the Slew Survey and $F_X \ge 4
\times 10^{-12} \rm erg \, cm^{-2} \, sec^{-1}$ in the EMSS we obtain
a sample 
that combines the advantage of a flux limited sample with a wide
coverage of the (spectral) parameter space and thus offers a representative subset of the BL Lac population.
This allows us to include several
"flavors" of BL Lac objects, ranging from the extreme X-ray dominated HBL to the borderline objects of intermediate BL Lac objects (IBL).  The objects we
analyse here are the second half of the sample, while the first 10
objects have been presented in Wolter et al. (\cite{wolter}),
hereafter W98.

\section{\sax Observations and Data Analysis}

The X-ray observations have been carried out with \sax, a project of
the Italian Space Agency (ASI) with a participation of the Netherlands
Agency for Aerospace (NIVR). A detailed description of the entire \sax
mission can be found in Butler \& Scarsi (\cite{sax}) and Boella et
al. (\cite{sax2}). The data presented here are from the Low Energy
Concentrator Spectrometer (LECS), the Medium Energy Concentrator
Spectrometer (MECS), and the Phoswich Detector System (PDS).

The LECS is sensitive in the 0.1--10 keV range with a field of view of
37 arcmin diameter (Parmar et al. \cite{lecs}). The LECS data are
useful up to 4 keV only, because the response matrix is not well
calibrated above this energy (Orr et al. \cite{orr}). It has an energy
resolution of a factor $\sim 2.4$ better than the {\em ROSAT PSPC}
(Brinkmann \cite{pspc}), while the effective area is between $\sim 6$
and 2 times lower at 0.28 and 1.5 keV, respectively.

The MECS has a field of view of 56 arcmin diameter, works in the energy range
1.3--10 keV with an energy resolution of $\sim 8\%$ and an angular
resolution of $\sim 0.7$ arcmin (FWHM) at 6 keV. The effective area at
6 keV is $155 \rm \, cm^2$ (Boella et al., \cite{mecs} ).

The PDS is sensitive in the 13--200 keV band. All photons within the
field of view are counted, with no spatial information
(f.o.v. diameter $\sim 1.3 \, \rm deg$ FWHM). Thus other sources in
the vicinity of the target can contaminate the measurement.

\begin{table}
\caption[]{The objects of the sample}
\begin{tabular}{llll}
Name & R.A. (2000.0) & Dec. & Redshift\\
\hline
\object{1ES 0145+138}  & 01 48 29.8 & +14 02 16  & 0.125\\
\object{1ES 0323+022}  & 03 26 13.9 & +02 25 15  & 0.147\\
\object{1ES 0507$-$040} & 05 09 38.2 & $-$04 00 46& 0.304\\
\object{1ES 0927+500}  & 09 30 37.6 & +49 50 26  & 0.186\\
\object{1ES 1028+511}  & 10 31 18.5 & +50 53 34  & 0.361\\
\object{1ES 1118+424}  & 11 20 48.0 & +42 12 10  & 0.124\\
\object{1ES 1255+244}  & 12 57 31.9 & +24 12 39  & 0.141\\
\object{1ES 1533+535}  & 15 35 00.7 & +53 20 38  & 0.890\\
\object{1ES 1544+820}  & 15 40 15.6 & +81 55 04  &  ? \\ 
\object{1ES 1553+113}  & 15 55 43.2 & +11 11 20  & 0.360\\
\object{1ES 1959+650}  & 20 00 00.0 & +65 08 56  & 0.047\\
\end{tabular}\\
\label{names}
\end{table}

The 11 objects presented here (for positions and redshifts see
Tab. \ref{names}) have been observed between May 1997 and February
1999. The data have been pre--processed at the \sax SDC (Science Data
Center) and retrieved through the SDC archive.  Table~\ref{journal}
shows the journal of observations, including exposure times and net
count rates for the LECS, MECS, and PDS detectors.

\begin{table*}
\caption[]{Journal of \sax observations}
\begin{tabular}{lrrrrrrr}
Name & obs. date & LECS & LECS & MECS & MECS & PDS & PDS\\
 & & exp. time [sec] & net counts & exp. time [sec] & net counts & exp. time [sec] & net counts\\
\hline
1ES 0145$+$138 & 30-31/12/97 & 10576 & $73.0 \pm 9.5$ & 12443 & $78.7 \pm 11.0$ & - & -\\  
1ES 0323$+$022 & 20/01/98 & 6093 & $201.6 \pm 14.5$ & 14408 & $607.4 \pm 27.2$ & 6845 & $256 \pm 561$\\  
1ES 0507$-$040 & 11-12/02/99 & 9116 & $441.6 \pm 21.5$ & 20689 & $1460.2 \pm 40.4$ & 9094 & $1515 \pm 633$\\
1ES 0927$+$500 & 25/11/98 & 8436 & $568.4 \pm 24.3$ & 22712 & $1967.3 \pm 46.5$ &10129 & $692 \pm 571$\\
1ES 1028$+$511 & 1-2/05/97 & 4552 & $737.9 \pm 28.1$ & 12622 & $2448.7 \pm 50.2$ & 9484 & $2763 \pm 718$\\
1ES 1118$+$424 & 1/5/97 & 6027 & $236.5 \pm 15.6$ & 9982 & $541.3 \pm 24.1$ & 8496 & $170 \pm 147$\\ 
1ES 1255$+$244 & 20/6/98 & 2484 & $297.1 \pm 17.4$ & 6910 & $1037.9 \pm 33.1$ & - & -\\
1ES 1533$+$535 & 13-14/02/99 & 8321 & $319.4 \pm 18.5$ & 26773 & $931.6 \pm 35.5$ & 4056 & $308 \pm 285$\\
1ES 1544$+$820 & 13/02/99 & 8043 & $170.7 \pm 13.6$ & 23249 & $510.8 \pm 26.9$ & 10414 & $780 \pm 363$\\
1ES 1553$+$113 & 5/02/98 & 4421 & $1179.5 \pm 34.5$ & 10618 & $2157.6 \pm 47.3$ & 4671 & $542 \pm 363$\\
1ES 1959$+$650 & 4-5/05/97 & 2252 & $423.2 \pm 20.7$ & 12389 & $3243.4 \pm 57.6$ & 7348 & $830 \pm 516$\\
\end{tabular}
\label{journal}
\end{table*}

\subsection{Spectral Analysis}

For observations accomplished in May 1997 all three MECS units are
available, while after this date only MECS units 2 and 3 were in use. 
The spectra from the MECS units have been summed together to increase
the S/N. None of the
sources shows extension neither in the LECS nor in the MECS image.
The same reduction process as in W98 has been applied to the data,
using FTOOLS v4.0 and XSPEC v.9.0 (Shafer et al. 1991).

We fitted simultaneously LECS and MECS data, leaving free the LECS
normalization with respect to the MECS to account for the residual
errors in the intensity cross-\linebreak calibration. A reliable value
for the ratio should be $LECS/MECS \sim 0.7 \pm 0.2$ (see the cookbook
for \sax scientific
analysis under
\small\verb+http://www.asdc.asi.it/bepposax/software/cookbook/+
\normalsize
and check the NFI flux cross-calibration).
For all sources where PDS data were available they were also 
simultaneously fitted: only for 1ES 1255+244 there are no PDS data and for
1ES 0145+138 the exposure time was too short to allow a detection in
the PDS. The PDS/MECS inter-calibration was handled as a free
  parameter. In all cases the consideration of the PDS part of the spectrum did not
  change significantly the results of the spectral analysis, also when fixing the
  inter-calibration  to the
  value of PDS/MECS $\sim 0.8$ as recommended by the \sax cookbook. In
  most cases the statistics of the PDS was not high, as can be seen
  from the net counts and errors given in Table \ref{journal}.

The assumed spectral model is a single-power law plus free low energy
absorption (arising from cold material with solar abundances; Morrison
\& McCammon \cite{morrison}).
For all objects we then checked whether a broken-power law with
  $N_H$ fixed to the Galactic value would give a
significantly better result (as measured by an F-test value greater than $95 \%$)
than the single-power law fit, and in five
cases indeed the fit improves. 

\begin{table*}
\caption[]{\sax: Best fit results for a single-power law model with free 
fitted $N_H$}
\begin{tabular}{lllllcrrccll}
Name & Energy & $N_H^{a}$ & $N_H^{a}$ & $F_X^{b}$ & $F_X^{c}$& Nm$^{d}$ & $\chi_{\nu}^2 (dof)$ & Prob. \\
 & Index $\alpha_X$& (Gal)&(Fit) & 2-10 & 0.5-2 &  &  & & & \\  
\hline
0145$+$138  & 1.50 $+0.94 \atop -0.68$ & 4.59 & 10.4 $+27.6 \atop -8.9 $& 0.35  & 0.45 & 0.86 & 0.38 (3)  & 77\%\\ 
0323$+$022  & 1.58 $+0.23 \atop -0.21$ & 7.27 & 31.6 $+14.6 \atop -11.5$& 2.24  & 1.93 & 0.81 & 0.51 (27) & 98\%\\ 
0507$-$040 & 1.14 $+0.12 \atop -0.11$ & 7.84 & 14.5 $+7.8 \atop -5.6  $& 3.88  & 2.72 & 0.82 & 0.73 (69) & 95\%\\ 
1028$+$511  & 1.32 $+0.08 \atop -0.07$ & 1.27 &  3.7 $+0.8 \atop -0.7  $& 10.10 &12.50 & 0.66 & 0.85 (97) & 85\%\\ 
1544$+$820  & 2.13 $+0.29 \atop -0.26$ & 3.70 & 20.1 $+11.3 \atop -8.2 $& 0.95  & 2.06 & 0.63 & 0.68 (30) & 90\%\\ 
1959$+$650  & 1.64 $+0.08 \atop -0.08$ & 0.99 & 25.5 $+7.1 \atop -6.0  $& 12.90 & 13.52& 0.65 & 0.79 (88) & 92\%\\ 
\end{tabular}

\label{powfreesax}
$^{a}$ hydrogen column density in $\rm \times 10^{20} cm^{-2}$\\
$^{b}$ un-absorbed flux in $10^{-12}\; \rm erg\,cm^{-2}\,sec^{-1}$ in
the 2--10 keV MECS energy band\\ $^{c}$ un-absorbed flux in $10^{-12}\;
\rm erg\,cm^{-2}\,sec^{-1}$ in the 0.5--2.0 keV LECS energy band\\
$^{d}$ Normalization of LECS versus MECS
\end{table*}

\begin{table*}
\caption[]{\sax: Best fit results for a broken-power law model
  applying Galactic $N_H$ values, when
significantly better representation of the data than the single
  power-law (according to an
F-test). For comparison, also the results for single-power law with
free fitted $N_H$ are given. The description of the various columns is 
as in Table~\ref{powfreesax}}
\begin{tabular}{lllllcrrccll}
Name & Energy & $\alpha_1$ & $\alpha_2$ & $E_0$ & $N_H^{a}$ & $N_H^{a}$ & $F_X^{b}$ & $F_X^{c}$& Nm$^{d}$ & $\chi_{\nu}^2 (dof)$ & Prob. \\
 & Index $\alpha_X$& & & [keV] & (Gal) & (Fit) & 2-10 & 0.5-2 & & \\  
\hline
0927$+$500  & 1.18 $+0.09 \atop -0.09$ & 0.40 $+0.18 \atop -0.23$ & 1.27 $+0.08 \atop -0.08$ & 1.35 $+0.28 \atop -0.24$ & 1.31 &  4.3 $+1.3 \atop -0.9  $& 4.59  & 4.59 & 0.69 & 0.88 (61) & 74\%\\
1118$+$424  & 1.57 $+0.16 \atop -0.16$  & 1.43 $+0.08 \atop -0.10$ & 3.43 $+0.7 \atop -0.7$ & 5.11 $+1.6 \atop -2.3$ & 2.59 &  3.5 $+1.3 \atop -1.0  $& 2.55  & 3.65 & 0.57 & 0.78 (23) & 76\%\\ 
1255$+$244  & 1.15 $+0.12 \atop -0.11$ & 0.61 $+0.19 \atop -0.37$ & 1.23 $+0.13 \atop -0.04$ & 1.58 $+0.36 \atop -0.36$ & 1.21 &  3.6 $+1.5 \atop -1.0  $& 8.03  & 7.81 & 0.74 & 0.78 (36) & 83\%\\ 
1533$+$535  & 1.57 $+0.15 \atop -0.14$ & 0.68 $+0.19 \atop -0.25$ & 1.74 $+0.16 \atop -0.15$ & 1.40 $+0.30 \atop -0.27$ & 1.28 &  4.8 $+1.9 \atop -1.1  $& 1.64  & 2.89 & 0.71 & 0.96 (44) & 55\%\\ 
1553$+$113  & 1.79 $+0.09 \atop -0.08$ & 0.57 $+0.18 \atop -0.22$ & 1.85 $+0.09 \atop -0.08$ & 1.13 $+0.90 \atop -0.90$ & 3.53 &  9.2 $+2.3 \atop -1.3  $& 9.37  & 19.16 & 0.76 & 1.05 (89) & 35\%\\ 
\end{tabular}
\label{brokenpowersax}
\end{table*}

We list the best fit parameters for the cases where the single power
law or broken-power law model gave the better fit in
Table~\ref{powfreesax} and Table~\ref{brokenpowersax}, respectively.
Galactic hydrogen column values were taken from the Leiden/Dwingeloo
Survey (Hartmann \& Burton \cite{LDS}) and are listed for
reference. The errors are 90\% confidence levels, for three interesting
parameters ($N_H$, $\alpha_X$, and normalization). Fluxes in the 2--10
keV band and also in the 0.5--2.0 keV band (for comparison with
ROSAT-PSPC fluxes) are given. Also listed are the normalization
factors of the LECS relative to the MECS.  For the broken-power law we
used the $LECS/MECS$ ratio as determined from the single-power law
fit and fixed $N_H$ to the Galactic value.

We then also checked if a broken-power law model with free fitted
absorption would give a better representation of the spectra. This
model increases the number of fit parameters and therefore the
complexity of the fit. Nonetheless we find all free fitted $N_H$
values lying in between the Galactic value and that from the
single-power law and only in one case (1ES 1959+650) the free $N_H$
from the broken-power law is in better agreement with the fitted $N_H$
from the single-power law than with the Galactic value.  Because of
that we assumed that in the broken-power law model there is no
significant low energy absorption in the source other than that
approximated by the Galactic neutral hydrogen column.
We thus have five free parameters in both models. For the single power
law with free fitted absorption the spectral slope $\alpha_X$, the
absorption $N_H$, the LECS/MECS and the PDS/MECS normalization,
and  normalization of MECS. Although for the 
broken-power-law we added two more free parameters, i.e. the break
energy $E_0$ and the second spectral index $\alpha_2$,
we fixed column density $N_H$ to the Galactic value and the
$LECS/MECS$ normalization to the same value found previously for the
single power law.
All broken-power laws show a flat slope in the low energy range
($\alpha_1 = 0.4 - 0.7$) and a steep high energy tail ($\alpha_2 =
1.2 - 1.9$) with a break energy within the LECS energy band ($E_0
= 1.1 - 1.6$ keV), except \mbox{1ES 1118+424} ($E_0 = 5.1$ keV).

\begin{figure*}
\epsfbox{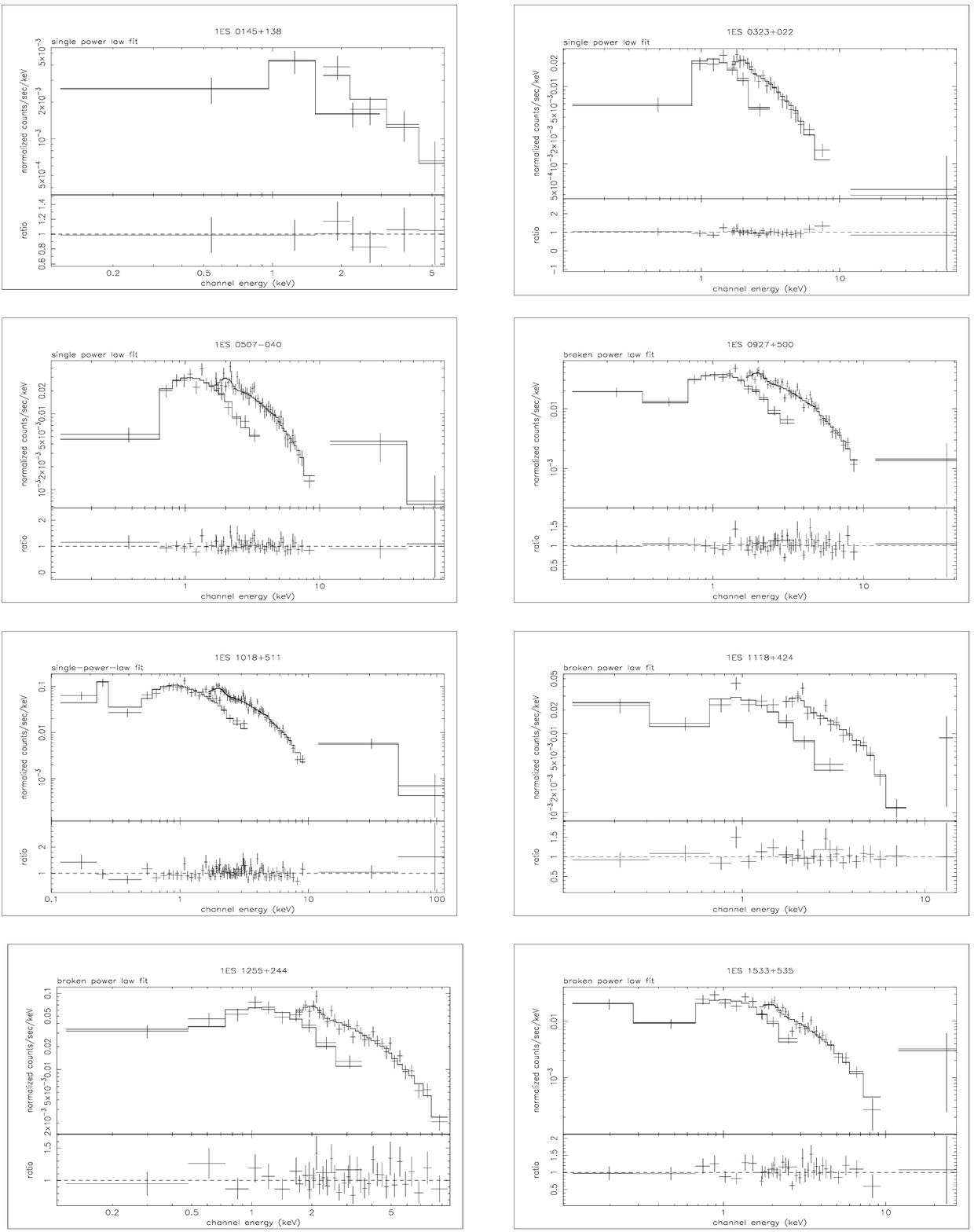}
\caption[]{\label{fig:fig1}Fits to the \sax spectra. In the five cases
where it gives a better spectral fit the result of the broken-power
law fit is shown.}
\end{figure*}

\addtocounter{figure}{-1} As mentioned, all sources except 1ES
1255+244 and 1ES 0145+138 have been also detected (at $\ge 4 \sigma$)
in the PDS. In all cases the single- or broken- power law fit give a
good approximation of the spectrum in the energy range $0.1 \leq E
\leq 30\, \rm keV$ and even up to 100 keV in the case of 1ES 1028+511.
       
The \sax data and best fits are plotted in Figure \ref{fig:fig1}.

Before concluding this section, we would like to point out the
presence of a possible feature in the SAX spectrum of 1ES
0927+500. For all models it shows a significant excess within the LECS
1.3 -- 1.5 keV energy range. For an added Gaussian line, the fit gives
a central energy $E = 1.4 \pm 0.1 \; \rm keV$, corresponding to a rest
frame energy of $\sim 1.7 \; \rm keV$, and a line width of $FWHM
\simeq 200 \; \rm eV$, comparable to the energy resolution of the LECS
at this energy (F-test probability $99.9\%$). Since the significance
of the detection is low and the energy does not have an obvious
physical identification, we refrain from further analysis until new
data will be available.

\begin{figure*}
\epsfbox{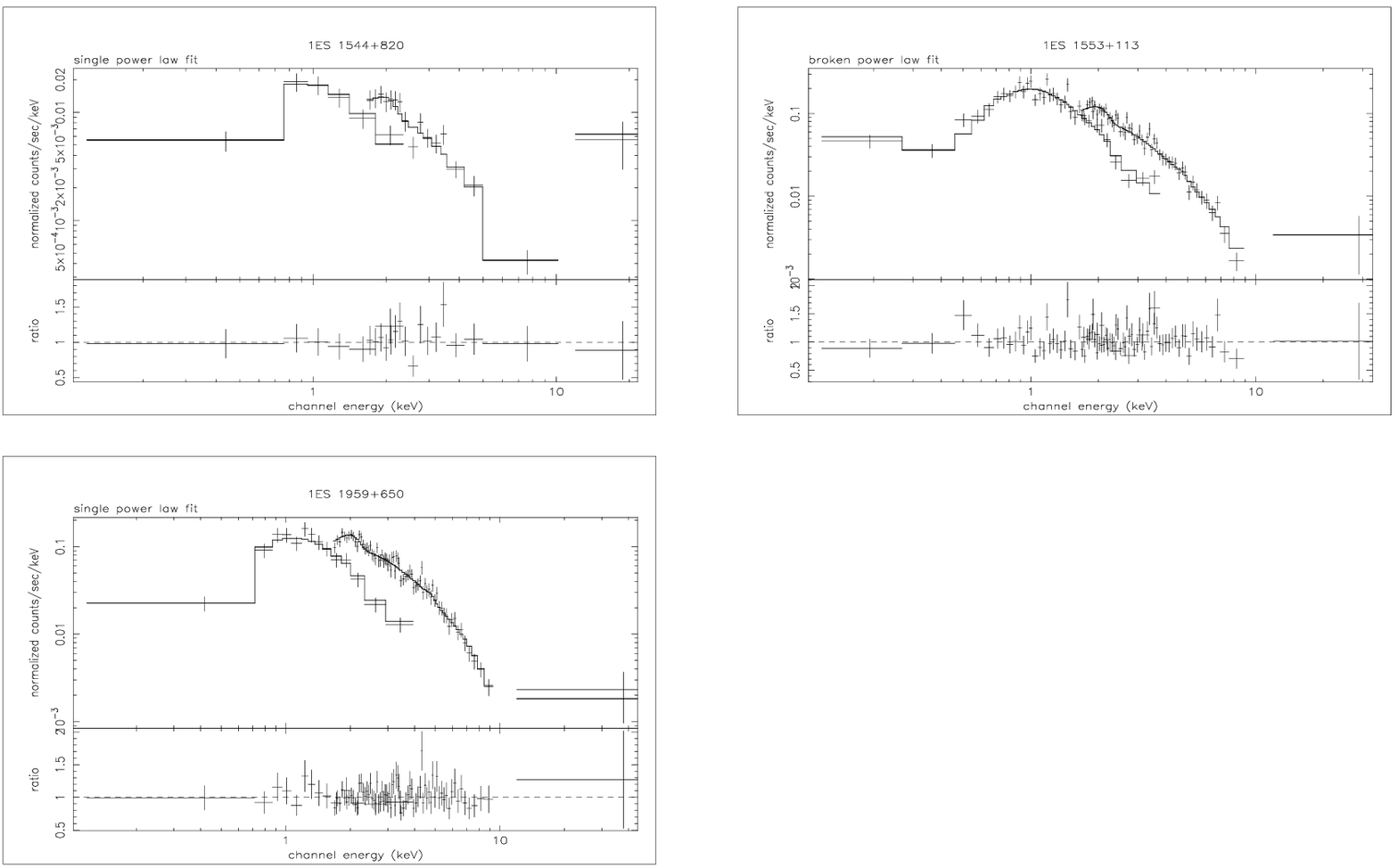}
\caption[]{\label{fig:fig2}(continued)} 
\end{figure*}

\section{Comparison with ROSAT PSPC data}

Four of the 11 sources studied here have also been observed during
{\em ROSAT PSPC} pointed observations and for ten of the
sources\footnote{The flux of 1ES 0145+138 is too low to be included
in the RASS-BSC and thus we do not have ROSAT data for this object.}
data are available from the ROSAT All Sky Survey Bright Source
Catalogue (RASS-BSC; Voges et al. \cite{bsc}).

In order to compare the ROSAT and \sax results we estimated the best
fitting values for spectral slope and absorption from the ROSAT
spectra. For the RASS-BSC data we used the hardness ratios, a method
described by Schartel et al. (\cite{schartel}), where the hardness
ratio is defined as $HR = \frac{H-S}{H+S}$ with $H$ and $S$ being the
number of counts in the hard and soft energy bands; typically two
ratios are computed: $HR1$ with energy ranges $S = 0.1 - 0.4 \rm \,
keV$ and $H = 0.5 - 2.0 \rm \, keV$, and $HR2$ with $S = 0.5 - 0.9 \rm
\, keV$ and $H = 1.0 - 2.0 \rm \, keV$ (Voges et al. \cite{bsc}). The
values for the hardness ratios range by definition from -1 for
extremely hard to +1 for very soft X-ray spectra. The error estimate
for the $N_H$ and $\alpha_X$ values is based on the hardness ratios
only, not on the photon spectrum itself. Therefore this method does
not give $\chi^2$ values, but is able to determine $68 \% (1
\sigma)$ errors. This is done by exploring the hardness-ratio,
spectral slope, and $N_H$ parameter space, determining the $1 \sigma$
region within it for a given set of parameter components. For the pointed
observations we instead used the standard reduction procedure as
described e.g. in Comastri, Molendi \& Ghisellini (\cite{pspcreduc}).

For all objects we considered two models, namely a single-power law
with either free fitted absorption or absorption fixed at the Galactic
value. For the pointed observations the model with free absorption
gives acceptable results; however the exposure time for the RASS-BSC
data is quite short and thus we used the values derived from the
single-power law model with Galactic absorption. The results are
reported in Tables~\ref{pspcpoint} and \ref{rass}, respectively.

\begin{table*}
\caption[]{ROSAT PSPC:
Results for the pointed observations for two models: a
single-power law with free fitted and with galactic absorption}
\label{pspcpoint}
\begin{tabular}{llrrcrrcrr}
Name & $F_X^{a}$ & $N_{H,Fit}$ & $F_{1\rm keV} [\mu Jy]$ &
$\alpha_{\rm ROSAT}^{b}$ & $\chi_{\nu}^2 (\rm d.o.f)$ & $N_{H,Gal}$ &
$\alpha_{\rm ROSAT}^{c}$ & $\chi_{\nu}^2 (\rm d.o.f.)$& exp.time [sec]\\
\hline
1ES 0323$+$022  & 4.97 & $9.7 \pm 0.3$ & 1.57 & $1.64 \pm 0.47$ & 2.83 (17) & 7.27 & $1.39 \pm 0.25$ & 7.31 (18)& 25304\\
1ES 0927$+$500  & 2.43 & $2.6 \pm 0.4$ & 0.80 & $1.39 \pm 0.13$ & 0.90 (17) & 1.31 & $0.92 \pm 0.06$ & 1.76 (18) & 3831\\ 
1ES 1028$+$511  & 7.78 & $1.4 \pm 0.7$ & 2.66 & $1.51 \pm 0.03$ & 3.46 (17) & 1.27 & $1.43 \pm 0.01$ & 3.65 (18)& 10777\\   
1ES 1118$+$424  & 1.54 & $2.6 \pm 0.3$ & 0.48 & $1.79 \pm 0.12$ & 1.16 (17) & 2.59 & $1.79 \pm 0.03$ & 1.09 (18)& 6416\\ 
\end{tabular}\\
$^{a}$ un-absorbed flux in the ``hard'' PSPC band (0.5--2.0 keV) in $10^{-12}\; \rm erg \; cm^{-2} sec^{-1}$\\
$^{b}$ spectral index for free fitted $N_H$\\
$^{c}$ spectral index for $N_H$ fixed to the Galactic value\\
\end{table*}

\begin{table*}
\caption[]{ROSAT PSPC: Results for the RASS data for two models: a
single-power law with free fitted and with galactic absorption}
\label{rass}
\begin{tabular}{llrrcrrrr}
Name & $F_X^{a}$ & $N_{H,Fit}$ & $F_{1\rm keV} [\mu Jy]$ & $\alpha_{\rm ROSAT}^{b}$ & $N_{H,Gal}$ & $\alpha_{\rm ROSAT}^{c}$ & net counts$^{d}$\\
\hline
1ES0323$+$022 & 12.3 & 18.9 & 1.49 & 2.58 $+1.67 \atop -1.03$ & 7.27 & 0.94  $+0.17 \atop -0.21$& 301.9 $\pm 24.2$\\
1ES0507$-$040 &  9.3 &  8.8 & 2.54 & 0.98 $+0.97 \atop -0.82$ & 7.84 & 0.82  $+0.23 \atop -0.27$& 274.0 $\pm 23.5$\\
1ES0927$+$500 & 18.1 &  2.2 & 2.57 & 1.20 $+0.32 \atop -0.31$ & 1.31 & 0.85  $+0.04 \atop -0.04$ & 525.5 $\pm 17.0$\\
1ES1028$+$511 & 18.0 &  1.5 & 4.52 & 1.39 $+0.23 \atop -0.22$ & 1.27 & 1.29  $+0.03 \atop -0.03$ & 853.9 $\pm 17.7$\\
1ES1118$+$424 &  4.7 &  3.3 & 0.80 & 1.91 $+0.65 \atop -0.62$ & 2.59 & 1.64  $+0.08 \atop -0.08$ & 139.0 $\pm 8.2$\\
1ES1255$+$244 &  4.8 &  3.8 & 0.34 & 2.07 $+0.58 \atop -0.55$ & 1.21 & 1.07  $+0.07 \atop -0.07$ & 187.2 $\pm 9.5$\\
1ES1533$+$535 &  8.0 &  2.0 & 1.82 & 1.23 $+0.23 \atop -0.23$ & 1.28 & 0.92  $+0.03 \atop -0.03$ & 781.1 $\pm 20.2$\\
1ES1544$+$820 &  4.0 &  5.3 & 1.26 & 1.77 $+0.51 \atop -0.50$ & 3.70 & 1.29  $+0.09 \atop -0.09$ & 252.4 $\pm 13.1$\\
1ES1553$+$113 & 12.0 &  0.7 & 3.64 & 0.16 $+0.82 \atop -0.40$ & 3.53 & 1.16  $+0.16 \atop -0.16$ & 510.5 $\pm 46.0$\\
1ES1959$+$650 & 27.9 & 15.8 & 1.23 & 1.76 $+0.44 \atop -0.21$ & 0.99 & -0.60 $+0.03 \atop -0.01$ & 3990.2 $\pm 61.1$\\
\end{tabular}\\
$^{a}$ un-absorbed flux in the ``hard'' PSPC band (0.5--2.0 keV) in $10^{-12}\;
\rm erg \; cm^{-2} sec^{-1}$\\ $^{b}$ spectral index for free fitted
$N_H$\\ $^{c}$ spectral index for $N_H$ fixed to the Galactic value\\
$^{d}$ PSPC 0.5--2.0 keV band\\ 
\end{table*}

\begin{figure}
\epsfysize=7.5cm
\epsfbox{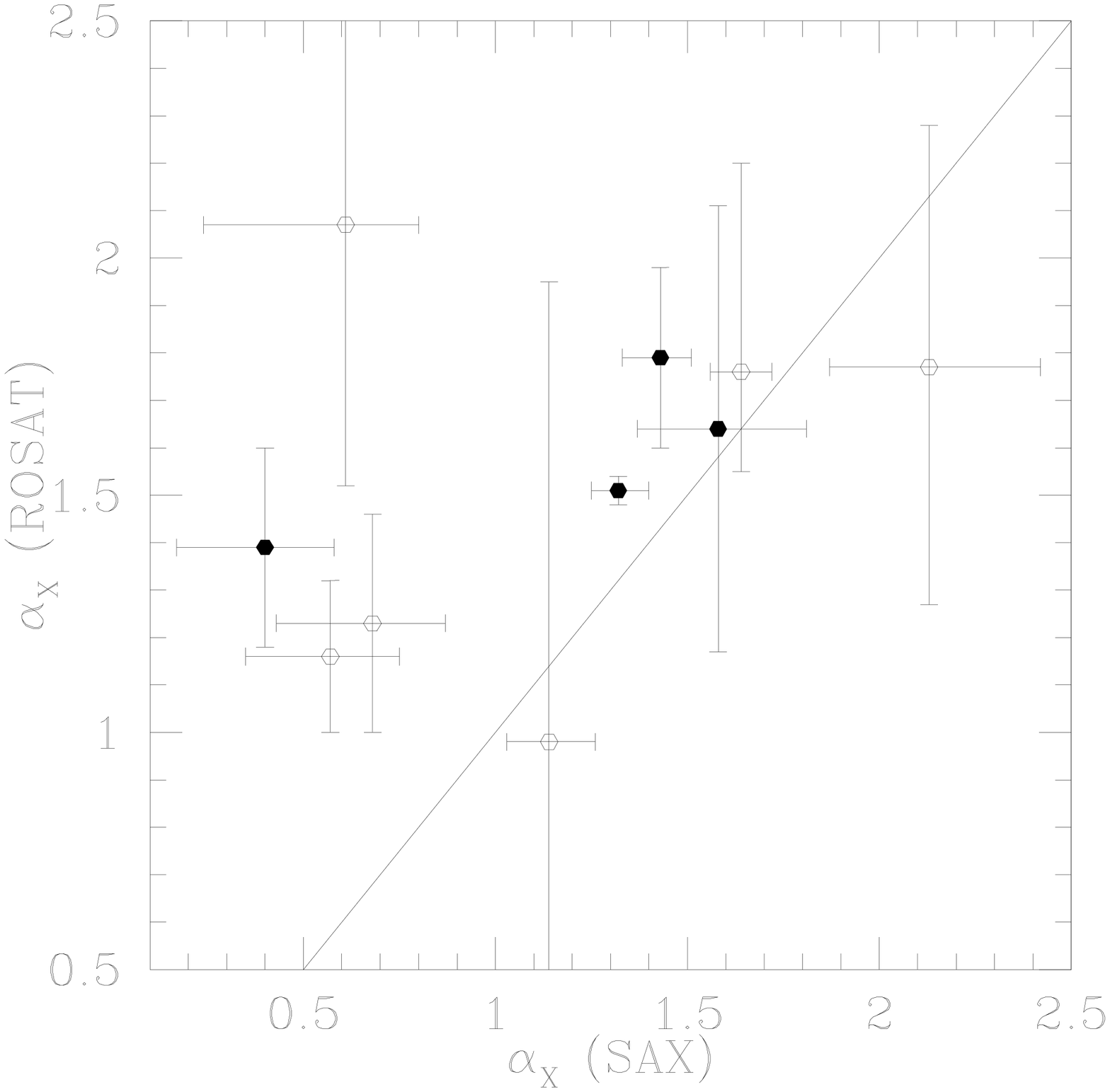}
\caption[]{\label{fig:alpha}X-ray spectral slopes derived from \sax in
comparison to those from {\em ROSAT PSPC} data. The dashed line represents
the $\alpha_{\sax} = \alpha_{ROSAT}$ relation. Open symbols refer to
the RASS-BSC data, while the others to {\em ROSAT PSPC} pointed
observations. For objects where the broken-power law fit to the \sax
spectra gave a better statistical result, the lower energy slope
$\alpha_1$ is reported.}
\end{figure}

\subsection{Comparison between \sax and ROSAT results}

We made a comparison of the fluxes between the {\em ROSAT PSPC} and
the \sax observations. To do so, we used the fluxes in the 0.5--2
  keV energy band, that is inside both the ROSAT/PSPC and LECS
energy ranges.
They vary by factors in the range $F_{\sax} / F_{\rm ROSAT} = 0.3 \dots 2.4$ 
with a mean ratio $F_{\sax} / F_{\rm ROSAT} = 1.1 \pm 0.7$. Actually,
high variability is expected in the X-ray flux of X-ray selected BL
Lac objects (e.g. Mujica et al. \cite{mujica}).  It is interesting
that for more radio powerful objects, like those belonging to the 1Jy
sample, this ratio is inverted, with a measured mean ratio $F_{\sax} /
F_{\rm ROSAT} = 0.6$ (Padovani et al. \cite{padovani01}).  
Radio selected BL Lac objects
have flat inverse Compton spectra extending toward high energies
(beyond $\sim 1 \rm \, keV$), but synchrotron emission, with a steep
spectrum, may often dominate at soft energies. Therefore, the fit of
their \sax LECS/MECS spectra, which sample a much wider energy range
toward the high energies than the ROSAT PSPC spectra do, may be
dominated by the flat inverse Compton component. Thus the fitted flux
at $1 \rm \, keV$ is lower than that determined with fits to ROSAT
spectra. This might cause the low $F_{\rm \sax}/F_{\rm ROSAT}$ ratio
computed by Padovani et al. This problem does not affect the hard and
flat soft X-ray spectra of our targets, so that our computed  $F_{\rm
  \sax}/F_{\rm ROSAT}$ ratio is consistent with unity.

The comparison between the ROSAT and \sax spectral slopes is shown in Figure \ref{fig:alpha}. Again we
used the spectral indices derived from the pointed PSPC observations
if available and the RASS-BSC spectral slopes otherwise. In fact the
best way would have been to fit the spectra in the same energy ranges
for both instruments, as we did for the flux values. But we were
prevented from doing so by the low statistics we would have had. The
spectral slope values
vary between the different observations, but the average difference is
${\alpha_{ROSAT} - \alpha_{\sax}} = 0.0 \pm 0.4$ (if we consider only
the objects with PSPC pointed observations we get ${\alpha_{ROSAT} -
\alpha_{\sax}} = 0.17 \pm 0.07$, while for the RASS-BSC data
\linebreak ${\alpha_{ROSAT} - \alpha_{\sax}} = -0.1 \pm 0.5$). While
these spectral indices are consistent with each other it is remarkable
that the broken-power law model applied to the \sax data shows in the
low energy band ($E < 1.5$ keV) a slope flatter than that of the {\em
ROSAT PSPC} (0.1 -- 2.4 keV).  The difference might be a hint of more
complex X-ray spectra or could result from the {\em ROSAT PSPC}
calibration problems discussed by Iwasawa et al. (\cite{iwa}) and
Barcons et al. (\cite{barcons}), who showed the X-ray spectral slopes
measured by the {\em ROSAT PSPC} to be significantly steeper than
those from other X-ray missions by $\Delta \alpha_X \sim 0.4$.  PSPC
spectra steeper than the \sax ones have also been found for the 1Jy
sample (Padovani et al. \cite{padovani01}).

\section{Spectral Energy Distribution}

To determine the broadband SED of the 11 objects we used radio data at
1.4 GHz from the VLA surveys NVSS (Condon et al. \cite{nvss}) or FIRST
(White et al. \cite{first}). Optical data were taken from the
literature and for some objects determined using the Calar Alto 1.23m
telescope (Beckmann \cite{photometry}). A campaign for simultaneous
optical data has also been performed for a few objects (Villata et al. 
\cite{magpaper}). 
Variability in the optical
band is not expected to be large: all objects presented here are X-ray
dominated objects (i.e. $\alpha_{OX} < 1.2$, see Table~\ref{sed})
which show only small optical variability (e.g. Villata et
al. \cite{magpaper}, Mujica et al. \cite{mujica}, Januzzi et
al. \cite{januzzi}).  With these data available we also computed the
broad band spectral indices (between $1.4 \rm GHz$, 4400 \AA\,  and 1
keV), which are reported in Table~\ref{sed}.

The following step consisted in parameterizing the derived SED. 

 Several authors pointed out that the synchrotron branch of the BL Lac
SED can be well approximated by a parabolic spectral shape (cf. Landau
et al. \cite{landau}, Comastri et al. \cite{pspcreduc}, Sambruna et
al. \cite{sambruna}, Fossati et al. \cite{fossati}, W98). We thus applied a
parabolic fit to the SED in the $\log \nu - \log \nu f_\nu$ plane --
parameterized as $\log \nu f_\nu = a \cdot (\log \nu)^2 + b \cdot \log
\nu + c$ -- and determined the peak position ($\nu_{peak}$) and the
total flux/luminosity of the synchrotron component.  The resulting
$\nu_{peak}$ and two-point indexes are
listed in Table~\ref{sed}. We also computed these quantities for the
objects reported in W98
\footnote{They slightly differ from the values presented there due to
the different reference energies: our $\alpha_{ox}$, $\alpha_{ro}$ and
$\alpha_{rx}$ are lower by $\sim 0.17$, $\sim 0.05$, and $\sim 0.09$
than those presented in W98. Furthermore after the publication of W98
two more redshifts for objects of the sample have been determined, 1ES
0502+675 ($z = 0.314$, Scarpa et al. \cite{hstsurvey}) and 1ES
1517+656 ($z = 0.702$, Beckmann et al. \cite{1517}). Thus we lack only
redshift information for 1ES 1544+820, for which we assume $z =
0.2$.}. 
In the following discussion the whole sample of 21 BL Lac
objects will be considered.

While the parabolic fits just described give a useful general
parameterization of the SED, we have also considered all the available
data and reproduced the SED of all sources of our sample with model
dependent representations.

\subsection{The blazar spectral sequence}

\begin{figure*}
\epsfysize=15cm
\epsfbox{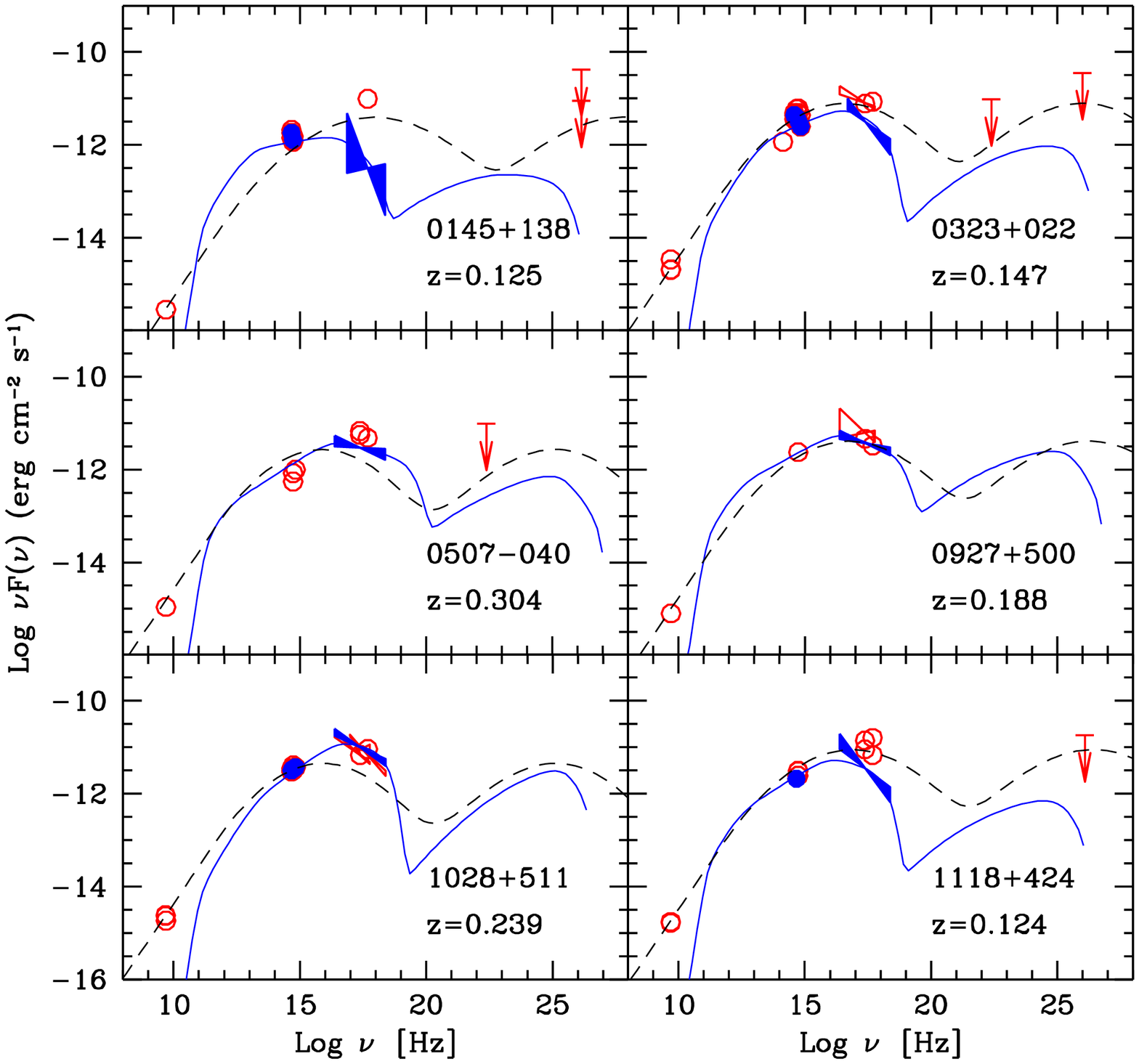}
\caption[]{\label{fig:sed1} Spectral energy distribution of our
sources. The solid lines correspond to the one--zone homogeneous
synchrotron and inverse Compton model calculated as explained in the
text, with the parameters listed in Table~\ref{params}.  The dashed
lines represent instead
the spectrum predicted by the phenomenological SED description by
Fossati et al. (\cite{fossati}) (with the changes proposed by Donato
et al. \cite{donato}).  The {\it Beppo}SAX data and simultaneous optical
data  of our observing campaigns
are indicated by the filled bow--ties and symbols.}
\end{figure*}

\addtocounter{figure}{-1} 
\begin{figure*}
\epsfysize=15cm
\epsfbox{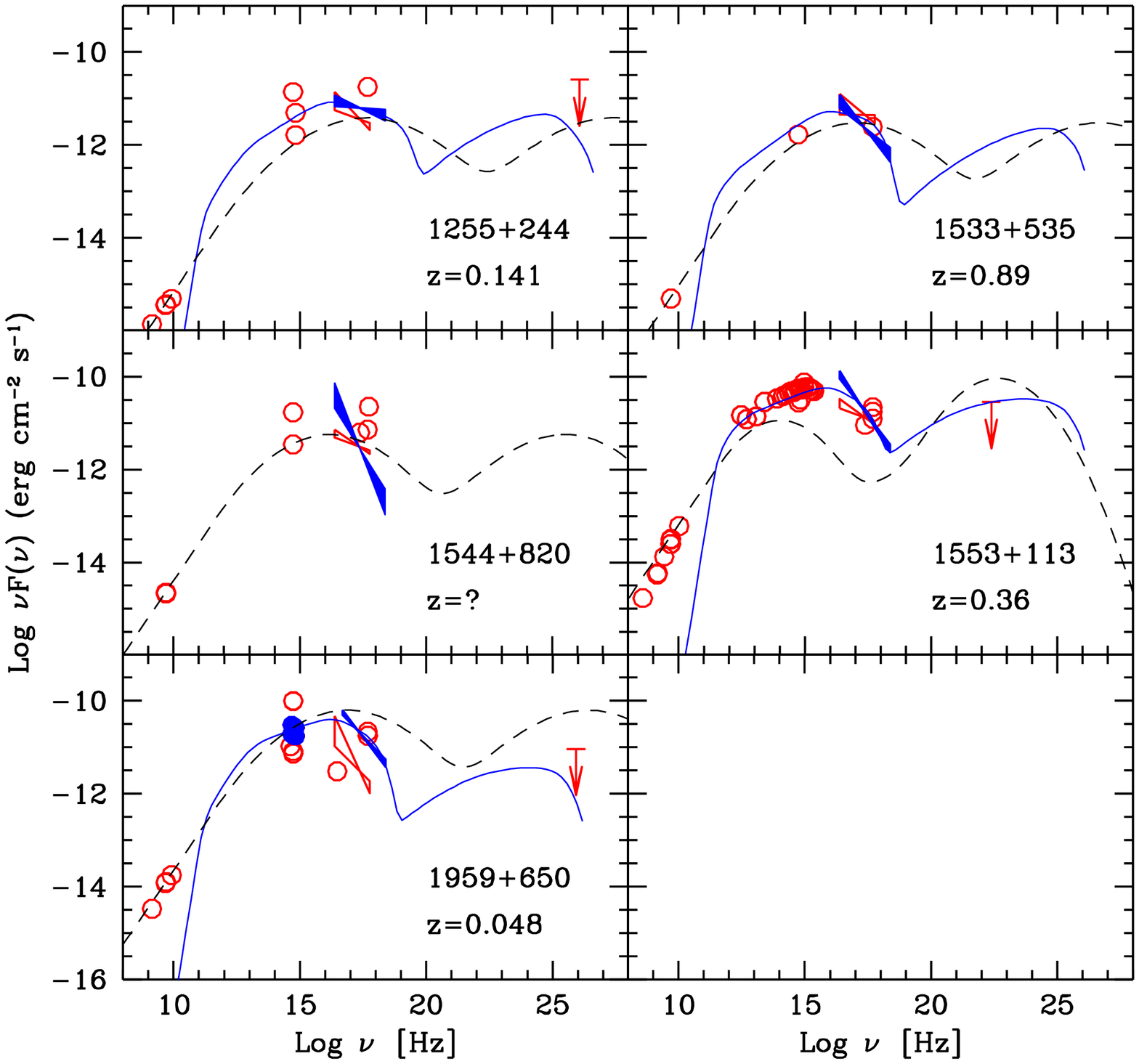}
\caption[]{\label{fig:sed2}(continued)}
\end{figure*}

Fossati et al. (\cite{fossati}) proposed a simple phenomenological
description of the average SED of blazars, based on the their
bolometric observed luminosity which is in turn assumed to be well
traced by the radio one.  The latter in fact was found to correlate
(and thus define) both the peak frequency of the synchrotron spectrum
and the relative importance of the inverse Compton vs synchrotron
powers.  More recently, Donato et al. (\cite{donato}) have revisited
the original parameterization by Fossati et al. (\cite{fossati}), but
(only) for objects below a radio luminosity of $10^{43}$ erg s$^{-1}$,
assuming a slightly different relation between the radio power, the
synchrotron peak frequency and the inverse Compton luminosity.  Thus
in the Donato et al. (\cite{donato}) model low power sources are
assumed to have comparable luminosities in the synchrotron and
self--Compton spectral components.  We have then applied this
parameterization to the objects in our sample, and the resulting `fits'
\footnote{Here not intended in a statistical sense, but only as
approximate representations of the SEDs} are shown as dashed lines in
Fig.~\ref{fig:sed1}. 

\subsection{Homogeneous Synchrotron Self--Compton models}

Finally we have reproduced the energy distributions adopting a more
`physical' approach by using an homogeneous (one-zone) synchrotron
self-Compton model (this is a ``one--zone version'' of the model
described in detail in Spada et al. \cite{spada}): it assumes that the
source is cylindrical, of size $R$ and width $\Delta
R^\prime=R/\Gamma$ (in the comoving frame, here $\Gamma$ is the bulk
Lorentz factor). 
The model is aimed at reproducing the spectrum originating in a limited 
part of the jet, the one thought to be responsible for most of the emission.  
This region is necessarily compact, since it must account for the 
fast variability shown by all blazars, especially at high frequencies.  
Therefore the radio emission from these compact sources is strongly 
self--absorbed, and the model cannot account for the observed radio flux.  

Sources in our sample are all BL Lacs of relatively low power with no
signs of strong broad emission lines in their spectra.  Consequently,
we have neglected the role of external seed photons for the inverse
Compton (IC) process.  

Relativistic electrons between a minimum and a maximum
energy are injected throughout the source
for a limited amount of time $t_{\rm inj}$
(mimicking a flare), which we set equal to $t_{\rm inj}=\Delta R^\prime/c$.
High energy electrons whose radiative loss timescales are shorter
than $t_{\rm inj}$ reach a stationary distribution, while
lower energy electrons retain the original spectrum.
The electron energies for which the cooling timescale is equal to
$t_{\rm inj}$ are denoted by $\gamma_{\rm c}$.
The particle distribution $N(\gamma)$ is assumed to have the slope $n$
[$N(\gamma)\propto \gamma^{-n}$] above $\gamma_{\rm c}$, 
while below this value there can be two cases.
We in fact assume that the particle distribution derives
from a continuous injection of particles between $\gamma_{\rm min}$ and
$\gamma_{\rm max}$, and the slope below $\gamma_{\rm c}$ depends on whether 
$\gamma_{\rm c}$ is greater or smaller than $\gamma_{\rm min}$.
If $\gamma_{\rm c}>\gamma_{\rm min}$, we have
$N(\gamma) \propto \gamma^{-n+1}$ between $\gamma_{\rm min}$
and $\gamma_{\rm c}$ and $N(\gamma)\propto \gamma^{-1}$ 
below $\gamma_{\rm min}$.
Alternatively, if $\gamma_{\rm c} < \gamma_{\rm min}$, 
then $N(\gamma) \propto \gamma^{-2}$ 
between $\gamma_{\rm c}$ and $\gamma_{\rm min}$.
We further assume that, below the minimum between $\gamma_{\rm c}$ and
$\gamma_{\rm min}$,
$N(\gamma)\propto \gamma^{-1}$.
According to these assumptions, the random Lorentz factor 
$\gamma_{\rm peak}$ of the electrons emitting most of the radiation 
(i.e. emitting at the peaks of the SEDs) becomes equal to
  $\gamma_{\rm c}$ and is determined by  
the importance of radiative losses and can assume values
within the range $\gamma_{\rm min}$--$\gamma_{\rm max}$.
Its value is listed in the last column of Table~\ref{params}.
The source is assumed to emit an intrinsic luminosity $L^\prime$
and is assumed to be observed with the viewing angle $\theta$.
All these input parameters are listed in Table~\ref{params}.

As can be
noted from it, the intrinsic luminosities, the source dimensions, the
bulk Lorentz factors and viewing angles, and the magnetic field values
are very similar for all sources, in agreement with the fact that all
objects belong to the same ``flavor'' of blazars.  The resulting fits
are shown in Fig.~\ref{fig:sed1} as solid lines.

Let us compare these representations of the SEDs and in particular the
results of the parameterization according to the Fossati et al
(\cite{fossati})'s scenario with those inferred from the homogeneous
SSC model. It turns out that the former tends to overestimate the
Compton $\gamma$--ray emission (by construction its power is never
less than the synchrotron one), but it agrees with the existing upper
limits in the GeV and TeV bands in all cases but 1ES 1553+113 and 1ES
1959+650.  We stress that both these parameterizations are not meant to
describe accurately the SED of specific sources, but only the average
SED of sources of equal total power.  Keeping this in mind, we can
consider the difference between the two theoretical spectra for each
source as a measure of the uncertainty associated with the `theoretical'
description of the SED.

\begin{table}
\caption[]{Derived quantities: Two-point broad band spectral
indices$^{a}$, X-ray luminosities and peak frequencies (for the 11
objects from this paper and for the 10 BL Lacs from W98).}
\begin{tabular}{llllll}
Name & $\alpha_{\rm ox}$ & $\alpha_{\rm ro}$ & $\alpha_{\rm rx}$ &
$\log L_X^{b}$ & $\log(\nu_{peak})$\\ 
\hline
1ES 0145$+$138   & 1.19 & 0.41 & 0.65 &  43.42 & 14.49\\ 
1ES 0323$+$022   & 1.07 & 0.34 & 0.57 &  44.38 & 14.89\\
1ES 0507$-$040   & 0.69 & 0.52 & 0.58 &  45.26 & 18.05\\
1ES 0927$+$500   & 0.81 & 0.36 & 0.50 &  44.89 & 16.34\\
1ES 1028$+$511   & 0.81 & 0.33 & 0.48 &  45.86 & 16.08\\
1ES 1118$+$424   & 0.91 & 0.32 & 0.51 &  44.30 & 15.49\\
1ES 1255$+$244   & 1.14 & 0.15 & 0.46 &  44.88 & 15.09\\
1ES 1533$+$535   & 0.77 & 0.37 & 0.51 &  46.05 & 15.77\\
1ES 1544$+$820$^{c}$ & 0.71 & 0.47 & 0.55 &  44.11 & 16.15\\
1ES 1553$+$113   & 0.81 & 0.43 & 0.56 &  45.89 & 15.58\\
1ES 1959$+$650   & 1.05 & 0.31 & 0.54 &  44.11 & 15.00\\[0.5cm]
\object{MS 0158.5+0019} & 0.72 & 0.39 & 0.50 &  45.12 & 16.95\\ 
\object{MS 0317.0+1834} & 0.64 & 0.42 & 0.49 &  45.11 & 18.51\\ 
\object{1ES 0347$-$121}  & 0.67 & 0.34 & 0.44 &  45.05 & 16.97\\ 
\object{1ES 0414+009}   & 0.76 & 0.42 & 0.53 &  45.59 & 16.25\\ 
\object{1ES 0502+675}   & 0.59 & 0.34 & 0.42 &  46.00 & 18.10\\ 
\object{MS 0737.9+7441} & 0.84 & 0.39 & 0.54 &  44.93 & 15.72\\ 
\object{1ES 1101$-$232}   & 0.72 & 0.39 & 0.49 &  45.83 & 17.37\\ 
\object{1ES 1133+704}   & 1.04 & 0.43 & 0.62 &  43.69 & 14.96\\ 
\object{MS 1312.1$-$4221} & 1.04 & 0.25 & 0.50 &  44.85 & 15.15\\ 
\object{1ES 1517+656}   & 0.75 & 0.32 & 0.47 &  46.55 & 16.17\\ 
\end{tabular}\\ 
\label{sed}                                
$^{a}$ derived from source fluxes at 1 keV, 4400 \AA, and 1.4 GHz, resp.\\
$^{b}$ luminosity in the 2 -- 10 keV band (MECS) in [erg/sec] \\
$^{c}$ assuming a redshift $z=0.2$\\
\end{table}

\begin{table*}
\caption[]{Input parameters of the homogeneous synchrotron self-Compton model}
\begin{center}
\begin{tabular}{lllllllrl}
\hline
Name  &$L^\prime$    &$R$  &$B$ &$\Gamma$ &$\theta$ &$n$  &$\gamma_{min}$ &$\gamma_{peak}$ \\
      & erg s$^{-1}$ &cm   & G  &         &         &     &     &        \\
\hline
0145$+$146  &6.0$\times 10^{40}$  &8.0$\times 10^{15}$ &1.0 &11     &5.0      &3.8  &600    &27000   \\
0323$+$022  &1.0$\times 10^{41}$  &1.0$\times 10^{16}$ &0.9 &11     &3.7      &3.5  &800    &27000   \\
0507$-$040  &2.5$\times 10^{41}$  &1.0$\times 10^{16}$ &1.2 &15     &3.5      &3.3  &1200   &20000  \\   
0927$+$500  &6.0$\times 10^{41}$  &1.5$\times 10^{16}$ &0.7 &14     &4.8      &3.4  &400    &30000   \\
1028$+$511  &2.7$\times 10^{41}$  &1.0$\times 10^{16}$ &1.0 &14     &3.0      &3.2  &800    &26000   \\
1118$+$424  &2.0$\times 10^{41}$  &1.0$\times 10^{16}$ &1.5 &15     &5.0      &3.3  &300    &13000   \\
1255$+$244  &6.5$\times 10^{41}$  &1.0$\times 10^{16}$ &1.0 &12     &5.0      &3.3  &500    &18000   \\
1533$+$534  &1.5$\times 10^{42}$  &2.0$\times 10^{16}$ &1.0 &15     &2.6      &3.2  &200    &12000   \\
1553$+$113  &2.5$\times 10^{42}$  &3.0$\times 10^{16}$ &0.7 &15     &2.5      &3.55 &300    &13000   \\
1959$+$650  &8.0$\times 10^{40}$  &1.0$\times 10^{16}$ &1.2 &13     &4.0      &3.6  &500    &19000       \\
\hline 
\end{tabular}
\end{center}
\label{params}
\end{table*}

\subsection{Derived parameters and correlations}

We can now consider possible trends in the properties of the SED and
examine their consistency with the predictions of the blazar unifying
scenario discussed above (Fossati et al \cite{fossati}, Ghisellini et
al \cite{ghisellini}).  

\begin{figure}
\epsfysize=7.5cm 
\epsfbox{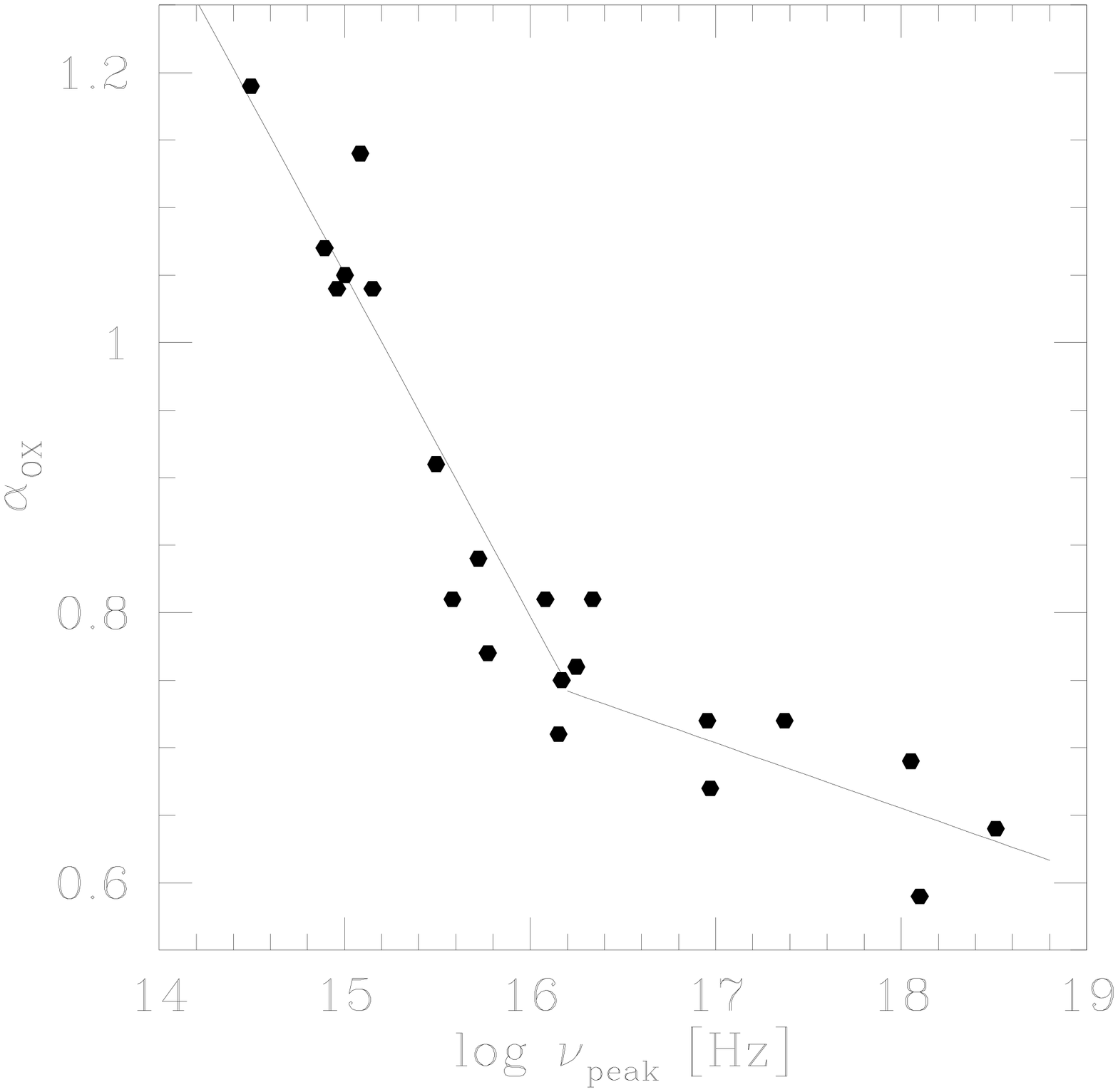}
\caption[]{\label{fig:aoxpeak}X-ray dominance ($\alpha_{ox}$) versus
the peak frequency of the synchrotron branch. The more X-ray dominated
objects have a higher peak frequency. The lines show two linear
regression fits with a break at $\log \nu_{peak} = 16.5$.}
\end{figure}

The first point we want to stress is that we find a strong correlation
between $\alpha_{ox}$ and the value of the peak frequency.  The
correlation is well represented by two linear regressions with a break
around $3 \times 10^{16} \, \rm Hz$ as shown in Figure
\ref{fig:aoxpeak}. (Note that small values of $\alpha_{ox}$ refer to
X-ray dominated objects while more optical dominated objects would
have $\alpha_{ox} > 1$.) Indeed this result is similar to that found
by Fossati et al. (\cite{fossati}) when studying the dependence of
$\alpha_{ro}$ and $\alpha_{rx}$ on the peak frequency.

\begin{figure}
\epsfysize=7.5cm
\epsfbox{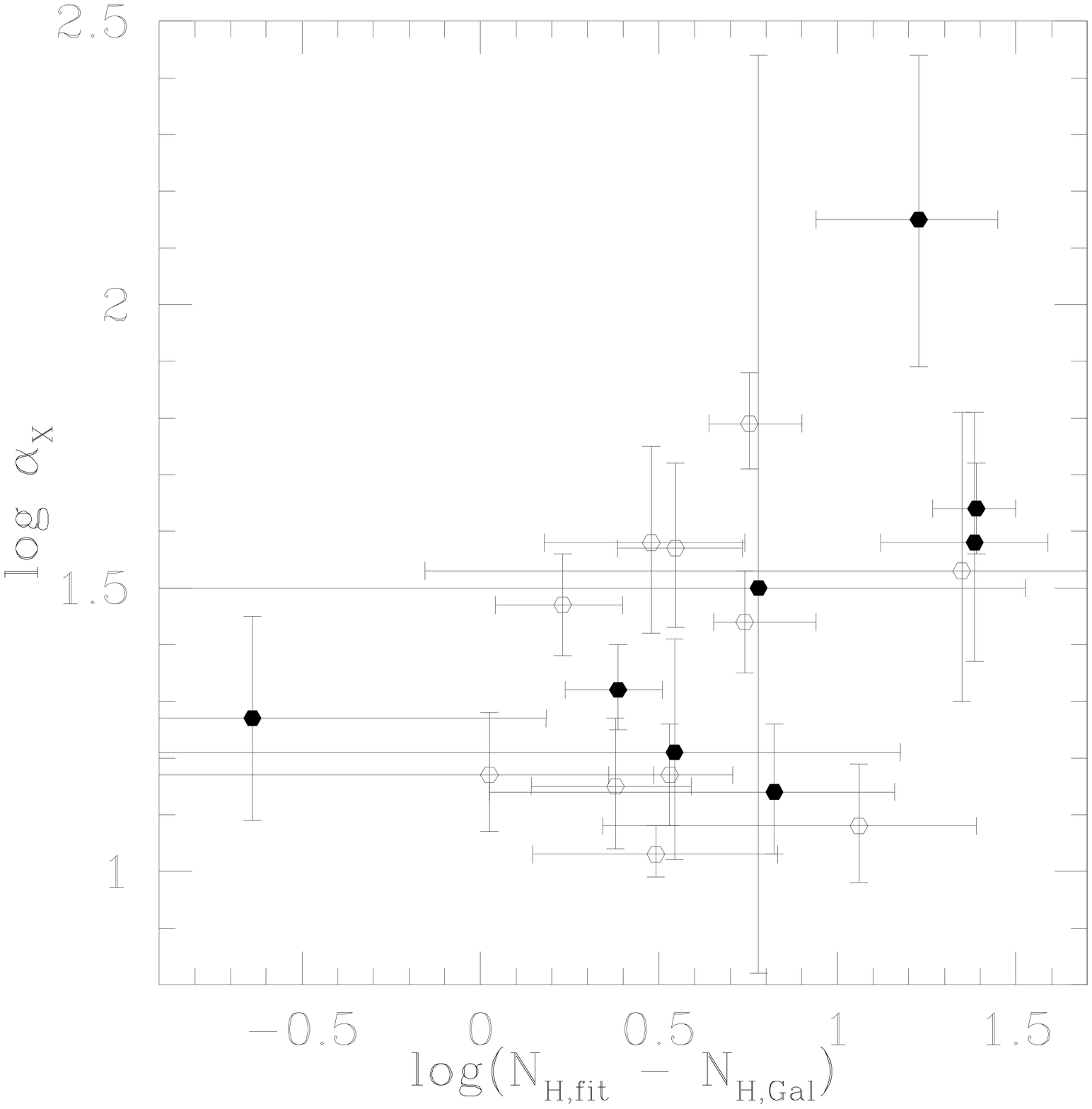}
\caption[]{\label{fig:nhax}Logarithmic ``intrinsic'' absorption
($N_{H,Fit} - N_{H,Gal}$ in $10^{20} \rm cm^{-2}$) versus X--ray
spectral slope. There is a trend for steeper spectra to correspond to
higher ``intrinsic'' absorption. Filled symbols represent the cases
where the single-power law fit was a better approximation to the \sax
spectra than the broken-power law. 
}
\end{figure}

\begin{figure}
\epsfysize=7.5cm
\epsfbox{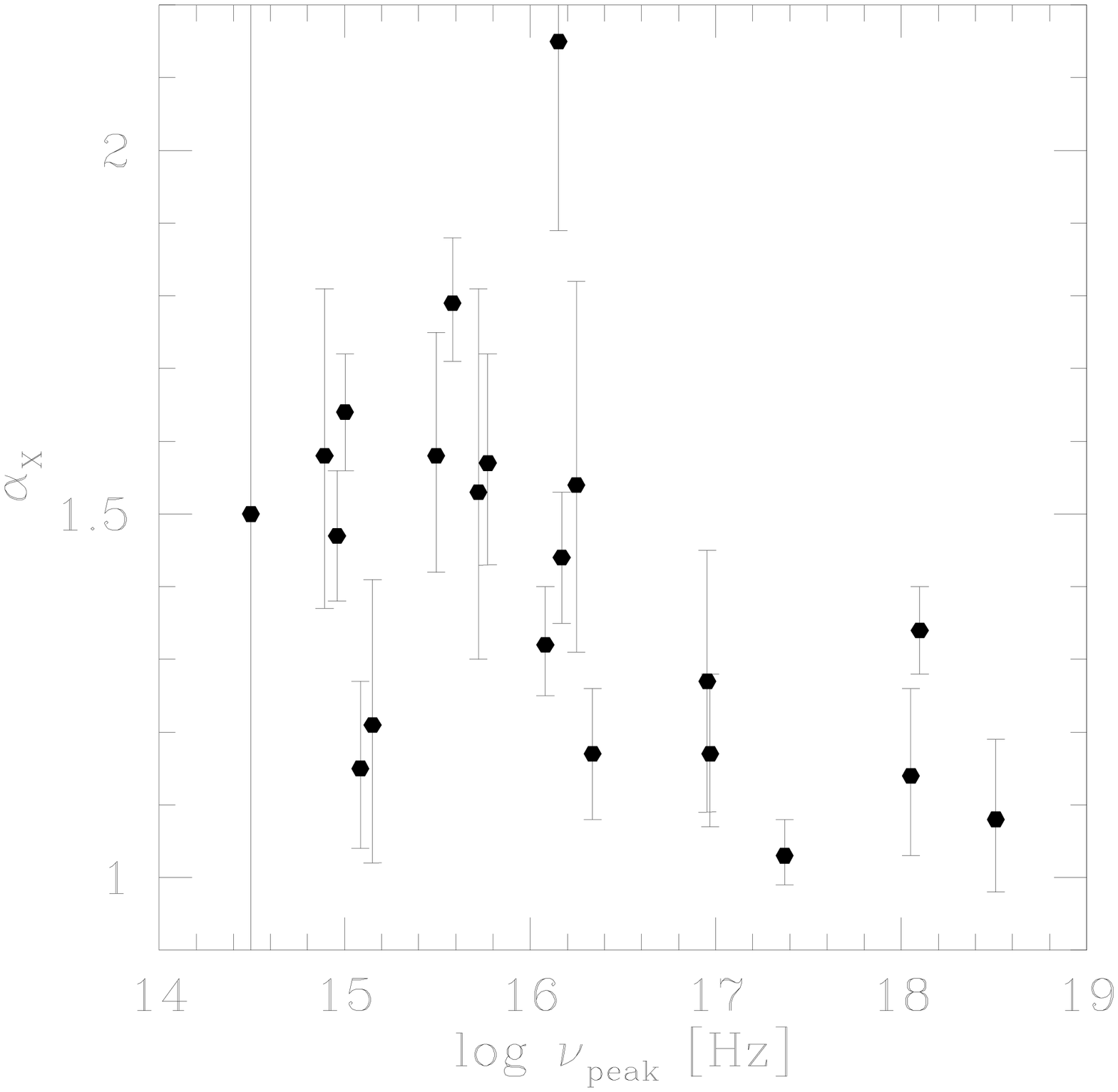}
\caption[]{\label{fig:peakax}X-ray spectral slope versus peak frequency.}
\end{figure}

A second interesting trend appears to be present between the intrinsic
absorption $\Delta N_H \equiv N_{H, Fit} - N_{H, Gal}$ and the X-ray
spectral slope $\alpha_X$.  For all fits the value of the absorbing
column was larger than the Galactic value. Even though error bars are
large and part of the correlation might be spurious, there is
indication of higher $\Delta N_H$ for steeper X-ray spectra (Figure
\ref{fig:nhax}).  If this trend were due to intrinsic absorption we
should also find a correlation with the X-ray luminosity, which
instead is not observed.
Alternatively, in a single power law model the absorption can
mimic an intrinsic curvature. Indeed a curvature of the X-ray spectra
can easily account for the $\Delta N_H$ vs $\alpha_X$ behavior: near
the synchrotron peak the spectra are harder and straighter than in the
steep decay at frequencies higher than the peak.
And in fact, the broken-power law model shows $N_H$ values in
agreement with the Galactic hydrogen column density (when the \sax
spectrum has enough counts to give a good statistics for this fit),
and in all cases the value for a free fitted $N_H$
for the pointed observations is in between the Galactic and the free
fitted value from the single-power law.  Furthermore we also find that
a higher peak frequency is associated with a flatter spectral slope
(Fig. \ref{fig:peakax}), a trend that again can be accounted for by
the above behavior: when the peak frequency is rising the X-ray band
is located near the maximum of the synchrotron branch and this gives
rise to flatter spectra.

\begin{figure}
\epsfysize=7.5cm
\epsfbox{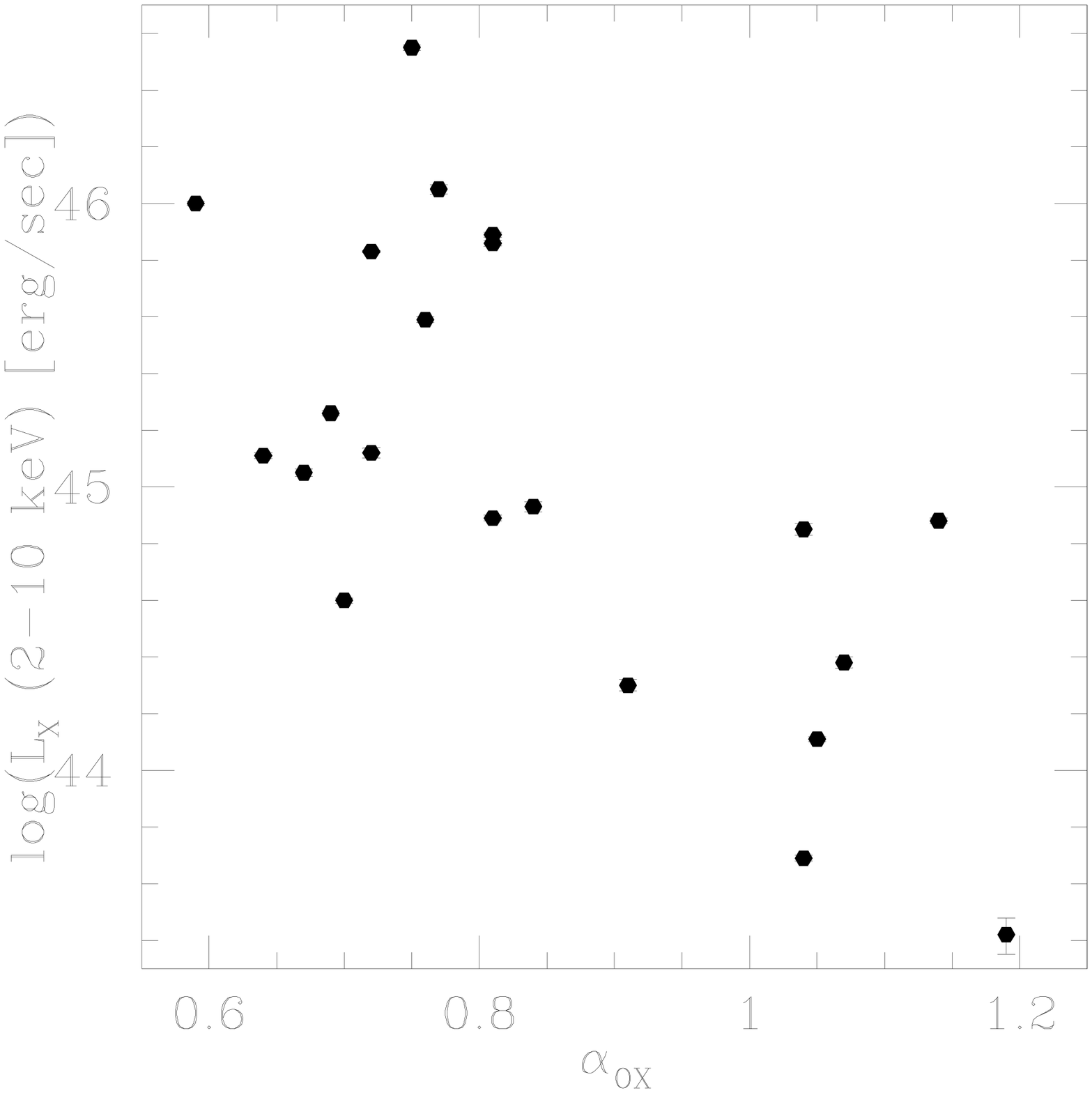}
\caption[]{\label{fig:aoxlx}X-ray luminosity for the \sax MECS band
(2--10 keV) versus X-ray dominance ($\alpha_{OX}$). Only for 1ES
0145+138 (bottom right) the error on the luminosity based on the count
rate statistics is larger than the plotted symbol.}
\end{figure}

A further correlation appears to be present between the X-ray
dominance and the X-ray luminosity (Figure \ref{fig:aoxlx}). Using a
linear regression we find that it can be described as $\log L_X = 47.7
- 3.2 \alpha_{OX}$ with a correlation coefficient of 0.7. This cannot
be due to selection effects (e.g. missing optical faint counterparts of
the weaker X-ray sources). In this case there should be also a correlation between the X-ray {\em fluxes} and the X-ray dominance, but this is not detectable. This trend was already reported by
Beckmann (\cite{beckmann}, \cite{thesis}) using a complete sample of
BL Lac objects based on the RASS-BSC.  

The same effect of X-ray
dominated objects appearing to have X-ray luminosities higher than the
intermediate BL Lacs should also be seen by comparing the peak frequency of
the synchrotron branch with the X-ray luminosity.
Within the scenario for the spectral properties of blazars discussed so
far it should also be expected that as the objects considered here have
similar radio power (i.e. total luminosity), their X--ray luminosity
is bound to be mostly affected by the location of the synchrotron
peak, higher energies corresponding to higher luminosities.
However, contrary to these predictions, there does not seem to be a
correlation between the peak frequency and the total luminosity of the
synchrotron branch (as determined through the parabolic fit to the
radio, optical and up to the X-ray data as described above). This
might indicate that when objects of the same class are considered the
range in total power is too small to show such a trend.  
The range of luminosities which is covered by the sample presented
here is indeed small: the radio luminosity ranges from $L_R = 1.2
\times 10^{24} \rm W/Hz$ to $L_R = 1.6 \times 10^{26} \rm W/Hz$, while
the spread in the X-ray luminosity is one decade larger ($1.4 \times
10^{19} \rm W/Hz < L_X < 1.7 \times  10^{22} \rm W/Hz$). For the
radio, optical, and X-ray energy band studied here, we find an average
of $\log \nu L_R = 41.1 \pm 0.6$, $\log \nu L_{opt} = 44.7 \pm 0.6$, and
$\log \nu L_X = 45.0 \pm 0.8$, which are consistent with the average SEDs of HBL as shown by Fossati et al. (\cite{fossati}). 
 
\section{Discussion and Conclusions}

The (physical) parameters of the objects studied here seem to partly 
correlate with the value of the peak frequency of the synchrotron
branch of the SED.  The X-ray selected sample considered in this work
comprises objects whose synchrotron peaks are located near/in the X-ray
band (high frequency peaked BL Lac, HBL; Padovani \& Giommi
\cite{padovani}). Those of radio selected BL Lac objects tend instead
to be located in the optical or infrared bands (low frequency peaked
BL Lac, LBL), as is the case of sources belonging to the 1-Jy BL Lac
sample (Stickel et al. \cite{stickel}) whose peak frequencies can be
as low as $\sim 10^{13} \rm Hz$ (Padovani et al. \cite{padovani01}).

We find a correlation between the peak frequency and the spectral
slope as well as between the spectral slope and the flattening of the
soft X-ray spectrum (parameterized as ``intrinsic'' absorption).  These
trends are consistent with the properties of the SED of HBLs: when the
peak of the SED moves into the X-ray region the X-ray spectra are
expected to become harder and straighter
as the peak frequency rises.  

This behavior also accounts for the relative high X-ray luminosities
observed in X-ray dominated BL Lacs (although LBL have higher total
bolometric luminosities than HBL their X-ray emission corresponds to
the energy band between the synchrotron and IC components,
e.g. Fossati et al. \cite{fossati}).  Although within our sample it
has not been possible to determine that HBL have lower bolometric
luminosity (as a large fraction of the bolometric luminosity is
expected to be emitted in the IC branch, which extends to energies
outside the frequency range we covered), we can assume - as already
mentioned - that the radio luminosity is a good tracer of the
bolometric one (Fossati et al. \cite{fossati}, Ghisellini
\cite{ghisellini99}).  
The lack of a visible trend between
such luminosities and the peak frequencies has then to be ascribed to
the small luminosity range spanned by the HBL sample in comparison with
the intrinsic dispersion of the peak-luminosity relation (as expected
e.g. for a distribution of the observing viewing angle which tends to
produce an opposite peak-luminosity trend).

If we consider this, the absence of correlation between the total luminosity
in the synchrotron branch and the peak frequency will then
result in higher bolometric luminosities for the LBL in comparison to
the HBL. This is consistent with the fact that LBL show strong IC components compared to the synchrotron branch, while HBL emit similar luminosities in the IC and the synchrotron branch. 

As the sample discussed here only includes HBL objects, we would expect
the X--rays to be dominated by the higher energy part of the
synchrotron and/or the lower energy part of the IC component
(Padovani et al. \cite{padovani01}, W98). Interestingly, no flattening
associated with the latter component is observed up to several tens of
keV (as the PDS data show that the X-ray power-law can be extended up
to $\sim 100 \rm keV$), implying that the IC seems to start emerging
at frequencies $\nu > 10^{19} \rm Hz$.  This is fully consistent with
the assumed SED for HBL as inferred from the sequence scenario
(Fossati et al. (\cite{fossati}).

In summary the \sax spectral survey shows that the X-ray properties of
X-ray selected BL Lac objects are in good global agreement with the
unified model for blazar, which ascribes the differences among blazars
mostly to the location of the peak frequencies of the synchrotron and
inverse Compton spectral components (e.g. Padovani \& Giommi
\cite{padovani}, Ghisellini et al. \cite{ghisellini}).  

Even though the correlation of peak frequency with total luminosity as described by Fossati et al. (\cite{fossati}) cannot be found in this sample, we find that the spectral properties are fully consistent with objects having a spread in the position of the peak of the synchrotron branch around the X-ray band. 
The sample as a whole represents only a part of the blazar population. Comparing the luminosities in the different energy bands studied here, the objects match the average SEDs of HBL as shown by Fossati et al. (\cite{fossati}).

The energy coverage provided by \sax toward the high energies
did not allow us to directly
examine the link between the position of the peaks and the bolometric
emitted power, as postulated by such model.  Such study will be
however possible with observations with sensitive instruments and good
spectroscopy resolution above $\sim 300$ keV as provided by e.g. the
INTEGRAL mission, which will cover the energy range of the spectral
change from the synchrotron to the IC dominated emission and possibly
of the peak of the IC component.

\begin{acknowledgements}
We thank the referee Elena Pian for constructive comments, which helped
to improve the manuscript. This research has made use of the NASA/IPAC Extragalactic
Database (NED) which is operated by the Jet Propulsion Laboratory,
California Institute of Technology under contract with the National
Aeronautics and Space Administration. We thank Team Members of the
\sax Science Operation Centre and Science Data Centre. The \sax
program is supported by the Italian Space Agency ASI.
VB thanks the 
{\it Osservatorio Astronomico di Brera} for the hospitality. 
This work has received partial financial support from the Deutsche
Akademische Austauschdienst (DAAD), the Gruppo Nazionale di Astronomia
of the CNR, and the Italian MURST.
\end{acknowledgements}


\begin{thebibliography}{}

\bibitem[1998]{bade}
Bade N., Beckmann V., Douglas N. G., et al., 1998, A\&A 334, 459
\bibitem[2000]{barcons}
Barcons X., Mateos S., Ceballos M. T., 2000, MNRAS 316, L13
\bibitem[1999]{beckmann}
Beckmann V., 1999, in: PASPC Vol. 159, eds. L. O. Takalo, A. Silanp\"a\"a
\bibitem[1999]{1517}
Beckmann V., Bade N., Wucknitz O., 1999, A\&A 352, 395
\bibitem[2000]{thesis}
Beckmann V., 2000, PhD thesis, Hamburg University, {\small\verb+http://www.sub.uni-hamburg.de/disse/330/vbdiss.html+}
\bibitem[2000a]{photometry}
Beckmann V., 2000a, BLAZAR Data 2, 3, {\small\verb+http://bldata.pg.infn.it/bldata.html+}
\bibitem[1997a]{sax2}
Boella G., Butler R. C., Perola G.C., et al., 1997, A\&AS 122, 299
\bibitem[1997b]{mecs}
Boella G., Chiappetti L., Conti G., et al., 1997b, A\&AS 122, 327
\bibitem[1991]{pspc}
Brinkmann W., 1991, In: Physics of Active Galactic Nuclei, Proceedings of the International Conference, Heidelberg, 3.-7.June 1991, Eds. Duschl, W. J., Wagner, S.J.
\bibitem[1990]{sax}
Butler C., Scarsi L., 1990, SPIE 1344, 46
\bibitem[1995]{pspcreduc}
Comastri A., Molendi S., and Ghisellini G., 1995, MNRAS 277, 297
\bibitem[1998]{nvss}
Condon J. J., Cotton W. D. Greisen E. W., et al.\ 1998, AJ 115, 1693
\bibitem[2001]{donato}
Donato D., Ghisellini G., Tagliaferri G., Fossati G., 2001, A\&A 375, 739
\bibitem[1998]{fossati}
Fossati G., Maraschi L., Celotti A, et al., 1998, MNRAS 299, 433
\bibitem[1998]{ghisellini}
Ghisellini G., Celotti A., Fossati G., et al., 1998, MNRAS 301, 451 
\bibitem[1999]{ghisellini99}
Ghisellini G., 1999, in PASPC Vol. 159, eds. L. O. Takalo,
A. Silanp\"a\"a, p. 311 
\bibitem[1990]{emss}
Gioia I. M., Maccacaro T., Schild R. E., et al., 1990, ApJS 72, 567
\bibitem[1997]{LDS}
Hartmann D., Burton W.B., 1997, ``Atlas of Neutral Galactic Hydrogen'', Cambridge University Press, Cambridge, New York
\bibitem[1999]{iwa}
Iwasawa K., Fabian A. C., Nandra K., 1999, MNRAS 307, 611
\bibitem[1994]{januzzi}
Januzzi, B. T., Smith, P. S., Elstan, R., 1994, ApJ 428, 130
\bibitem[1990]{polarization}
K\"uhr H., Schmidt G. D., 1990, AJ 99, 1
\bibitem[1986]{landau}
Landau R., Golisch B., Jones T. J., et al., 1986, ApJ 308, 78
\bibitem[1983]{morrison}
Morrison R., Mc Cammon D., 1983, ApJ 270, 119
\bibitem[1999]{mujica}
Mujica R., Appenzeller F.-J., Krautter I., et al., 1999, in: PASPC Vol. 159, eds. L. O. Takalo, A. Silanp\"a\"a
\bibitem[1998]{orr}
Orr A., Parmar A. N., Yaqoob T., Guainazzi M., 1998, in: Proceedings of the Active X-ray Sky symposium, October 21-24, 1997, Rome, eds. Scarsi et al.
\bibitem[2001]{padovani01}
Padovani P., Costamante L., Giommi P., et al., 2001, MNRAS 328, 931
\bibitem[1996]{padovani}
Padovani P., Giommi P., 1996, MNRAS 279, 526
\bibitem[1997]{lecs}
Parmar A.N., Martin D.D.E., Bavdaz M., et al., 1997, A\&AS 122, 309
\bibitem[1996]{ess}
Perlman E. S., Stocke J. T., Schachter J. F, et al., 1996,  ApJS 104, 251
\bibitem[1996]{sambruna}
Sambruna R. M., Maraschi L., Urry C. M., 1996, ApJ 463, 444
\bibitem[1999]{hstsurvey}
Scarpa R., Urry C. M., Falomo R., et al., 1999, ApJ 521, 134 
\bibitem[2000]{scarpa}
Scarpa R., Urry C. M., Falomo R., et al., 2000, ApJ 532, 74
\bibitem[1996]{schartel}
Schartel N., Walter R., Fink H. H., Tr\"umper J., 1996, A\&A 307,33
\bibitem[2001]{spada}
Spada M., Ghisellini G., Lazzati D., Celotti A., 2001, MNRAS 325, 1559
\bibitem[1991]{stickel}
Stickel M., Padovani P., Urry C. M., et al., 1991, ApJ 374, 431
\bibitem[1991]{stocke}
Stocke J. T., Morris S. L., Gioia I. M., et al., 1991, ApJS 76, 813
\bibitem[2000]{magpaper}
Villata M., Raiteri C. M., Popescu M. D., et al., 2000, A\&AS 144, 481
\bibitem[1992]{voges}
Voges W., 1992, in: Proceedings on European International Space Year Meeting ESA ISY-3, 9
\bibitem[1999]{bsc}
Voges W., Aschenbach B., Boller Th., et al., 1999, A\&A 349, 389
\bibitem[1995]{variability}
Wagner S. J., Witzel A., 1995, ARA\&A 33, 163
\bibitem[1997]{first}
White R. L., Becker R. H., Helfand D. J., Gregg M. D., 1997, ApJ 475, 479
\bibitem[1998]{wolter}
Wolter A., Comastri A., Ghisellini G., et al., 1998, A\&A 335, 899 (W98)
\end{thebibliography}
\end{document}